\documentclass[final]{aipproc}

\layoutstyle{6x9}
\setcitestyle{numbers}

\begin{document}
\begin{figure}[!ht]
\includegraphics*[height=.12\textheight]{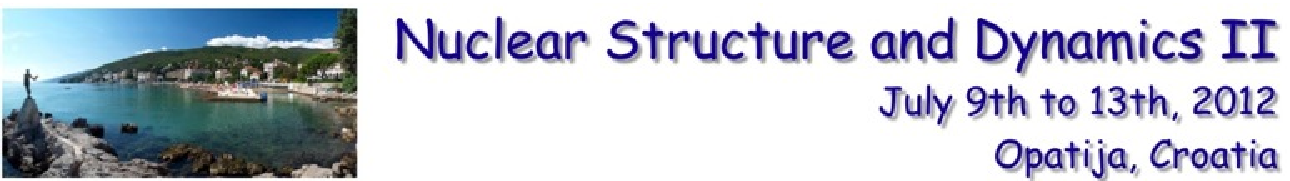}
\end{figure}

\title{Dynamic Microscopic Theory of Fusion Using DC-TDHF}


\author{A.S. Umar}{
  address={Department of Physics and Astronomy, Vanderbilt University, Nashville, Tennessee 37235, USA}
}
\author{V.E. Oberacker}{
  address={Department of Physics and Astronomy, Vanderbilt University, Nashville, Tennessee 37235, USA}
}
\author{R. Keser}{
  address={RTE University, Science and Arts Faculty, Department of Physics, 53100, Rize, TURKEY}
}
\author{J.A. Maruhn}{
  address={Institut f\"ur Theoretische Physik, Goethe-Universit\"at, D-60438 Frankfurt am Main, Germany}
}
\author{P.-G. Reinhard}{
	       address={Institut fur Theoretische Physik, Universitat Erlangen, D-91054 Erlangen, Germany}
}

\begin{abstract}
\noindent The density-constrained time-dependent Hartree-Fock (DC-TDHF) theory is a fully microscopic
approach for calculating heavy-ion interaction potentials and
fusion cross sections below and above the fusion barrier.
We discuss recent applications of DC-TDHF method
to fusion  of light and heavy systems.
\end{abstract}

\maketitle

\section{Introduction}

The investigation of internuclear potentials for heavy-ion collisions is
of fundamental importance for the study of fusion reactions
as well as for the formation of
superheavy elements and nuclei far from stability.
Recently, we have developed a new method to extract ion-ion interaction potentials directly from
the time-dependent Hartree-Fock (TDHF) time-evolution of the nuclear system~\cite{DC-TDHF1}.
In the density-constrained TDHF (DC-TDHF) approach
the TDHF time-evolution takes place with no restrictions.
At certain times during the evolution the instantaneous density is used to
perform a static Hartree-Fock minimization while holding the neutron and proton densities constrained
to be the corresponding instantaneous TDHF densities. In essence, this provides us with the
TDHF dynamical path in relation to the multi-dimensional static energy surface
of the combined nuclear system. Since TDHF directly provides us with the most probable fusion path
in the mean-field limit there is no need calculate the entire PES to determine the fusion path as
in the case of computation of fission barriers.
In short, we have a self-organizing system which selects
its evolutionary path by itself following the microscopic dynamics.
Some of the effects naturally included in the DC-TDHF calculations are: neck formation, mass exchange,
internal excitations, deformation effects to all order, as well as the effect of nuclear alignment
for deformed systems.
The DC-TDHF theory provides a comprehensive approach to calculating fusion barriers
in the mean-field limit. The theory has been applied to calculate fusion cross-sections
for large number of systems~\cite{DC-TDHF1,DC-TDHF2}.
In this proceeding we will outline the DC-TDHF method and give examples of its
application to the calculation of fusion cross-sections for various systems.

\section{Theory}
In the DC-TDHF approach, the time-evolution takes place with no restrictions.
At certain times $t$ or, equivalently, at certain internuclear distances
$R(t)$ the instantaneous TDHF density
\begin{equation}
\rho_{\mathrm{TDHF}}(r,t) = <\Phi(t) |\rho| \Phi(t) >
\label{eq:rho_TDHF}
\end{equation}
is used to perform a static Hartree-Fock energy minimization
\begin{equation}
\delta <\Phi_{\rho} \ | H - \int d^3r \ \lambda(r) \ \rho(r) \ | \Phi_{\rho} > = 0
\label{eq:var_dens}
\end{equation}
while constraining the proton and neutron densities to be equal to the instantaneous
TDHF densities
\begin{equation}
<\Phi_{\rho} |\rho| \Phi_{\rho} > = \rho_{\mathrm{TDHF}}(r,t) \ .
\label{eq:dens_constr}
\end{equation}
These equations determine the state vector $\Phi_{\rho}$. This means we
allow the single-particle wave functions to rearrange themselves in such a way
that the total energy is minimized, subject to the TDHF density constraint.
In a typical DC-TDHF run, we utilize a few
thousand time steps, and the density constraint is applied every $20$ time steps.
We refer to the minimized energy as the ``density constrained energy'' $E_{\mathrm{DC}}(R)$
\begin{equation}
E_{\mathrm{DC}}(R) = <\Phi_{\rho} | H | \Phi_{\rho} > \ .
\label{eq:EDC}
\end{equation}
The ion-ion interaction potential $V(R)$ is essentially the same as $E_{\mathrm{DC}}(R)$,
except that it is renormalized by subtracting the constant binding energies
$E_{\mathrm{A_{1}}}$ and $E_{\mathrm{A_{2}}}$ of the two individual nuclei
\begin{equation}
V(R)=E_{\mathrm{DC}}(R)-E_{\mathrm{A_{1}}}-E_{\mathrm{A_{2}}}\ .
\label{eq:vr}
\end{equation}
The interaction potentials calculated with the DC-TDHF method incorporate
all of the dynamical entrance channel effects present in the mean-field dynamics, such as neck formation,
particle exchange, internal excitations, and deformation effects.
While the outer part of the potential barrier is largely determined by
the entrance channel properties, the inner part of the potential barrier
is strongly sensitive to dynamical
effects such as particle transfer and neck formation.

Using TDHF dynamics, it is also possible to compute the corresponding coordinate
dependent mass parameter $M(R)$ using energy conservation at zero impact parameter.
As expected, at large distance $R$ the mass $M(R)$ is equal to the
reduced mass $\mu$ of the system. In general, we observe that the coordinate-dependent mass changes only
the interior region of the potential barriers, and this change is most pronounced
at low $E_\mathrm{c.m.}$ energies. Fusion cross-sections are calculated
by numerically integrating the Schr\"odinger equation using the well-established
{\it Incoming Wave Boundary Condition} (IWBC) method~\cite{HR99}.

\section{Results}
\begin{figure}[!htb]
\includegraphics*[height=.28\textheight]{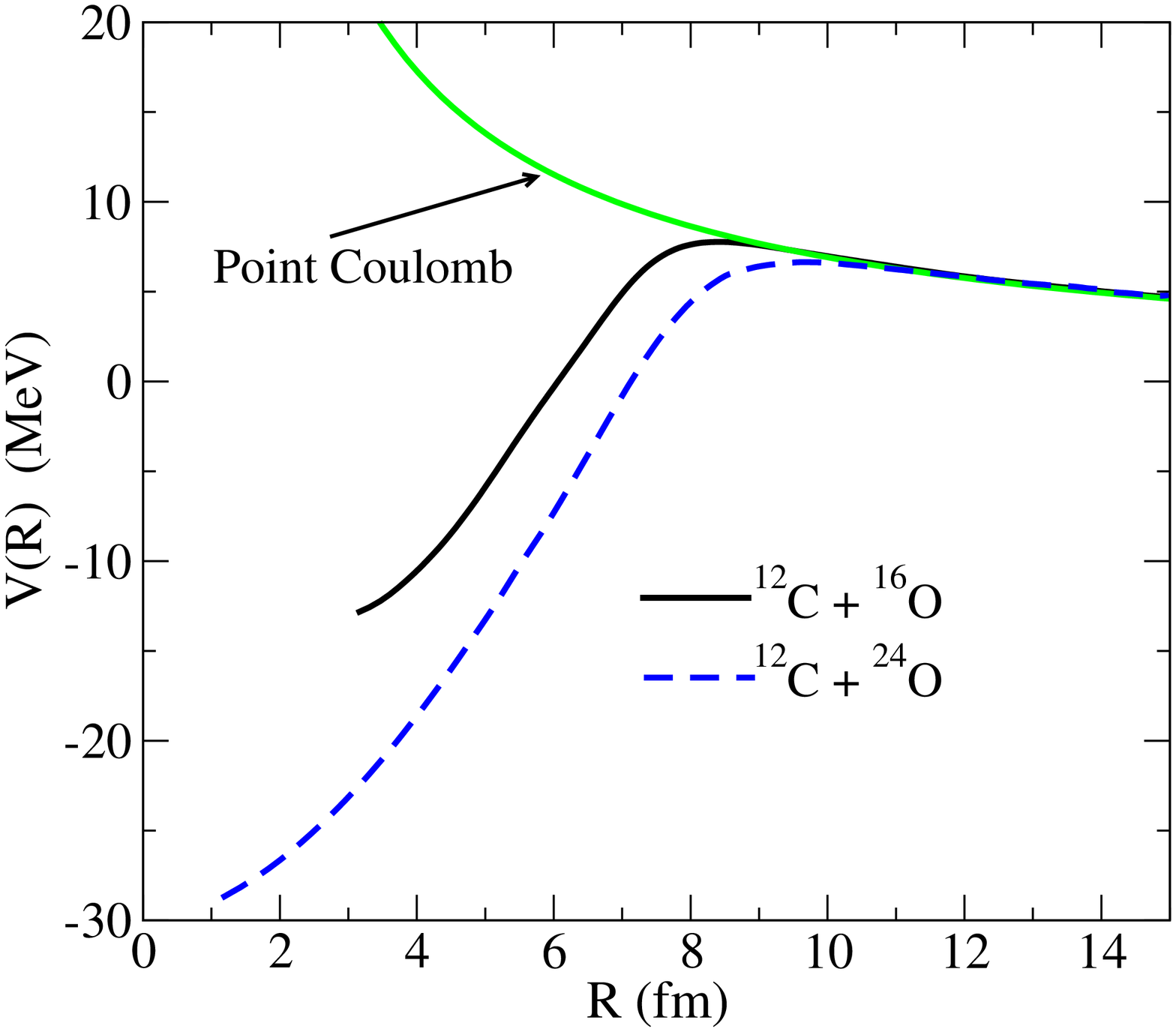}\hspace{0.04in}\includegraphics*[height=.28\textheight]{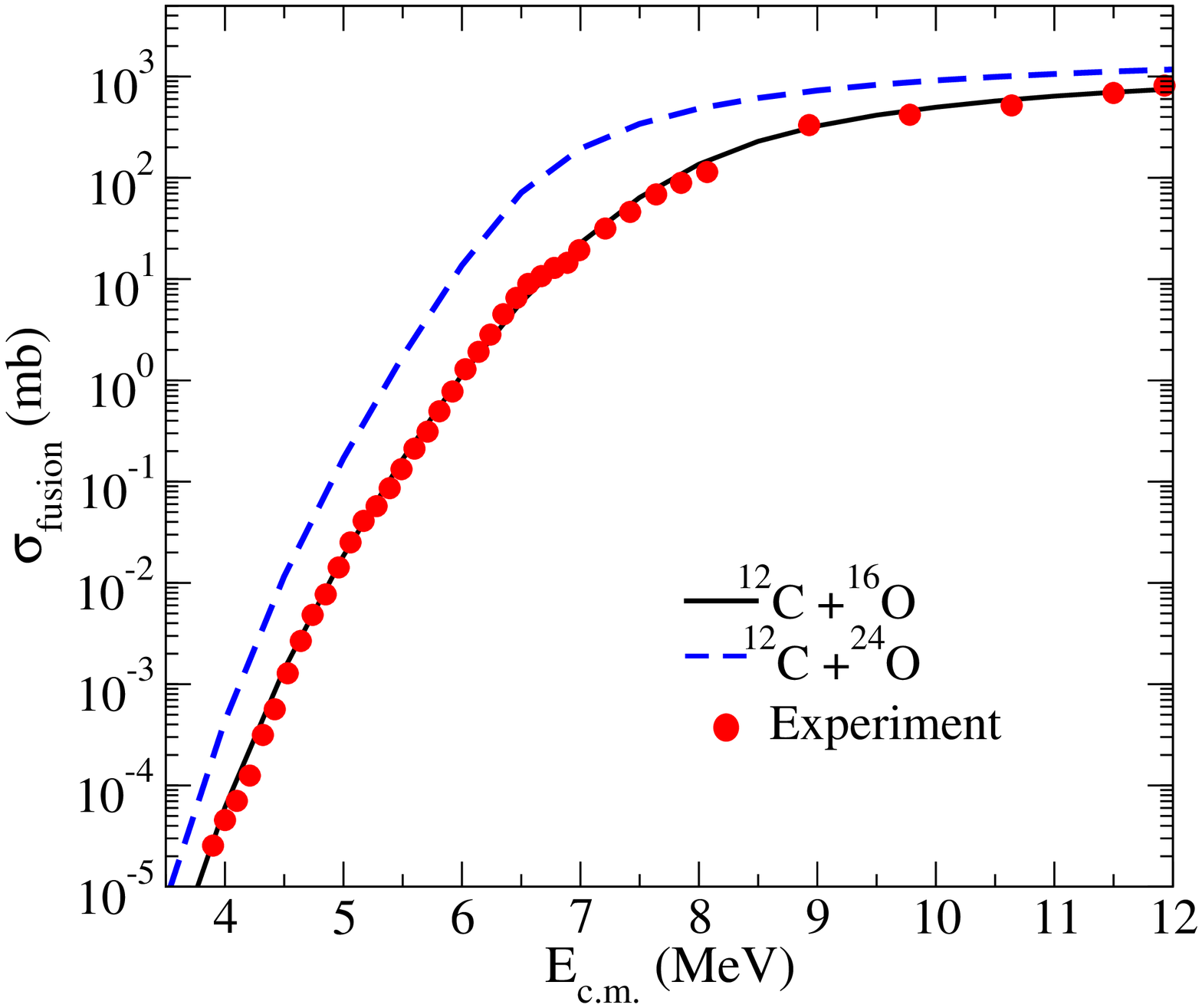}
\caption{\label{fig1} (a) Ion-Ion potential for various isotopes of the C$+$O system.
                      (b) Corresponding cross-sections.}
\end{figure}

In this Section we give some recent examples of DC-TDHF calculations of heavy-ion potentials and cross-sections.
Fusion of very neutron rich nuclei may be important to determine the composition and heating of the crust of accreting
neutron stars~\cite{DC-TDHF-OO}. In Fig.~\ref{fig1}a we show the DC-TDHF potential barriers for the C$+$O system.
The higher barrier corresponds to the $^{12}$C$+$ $^{16}$O system and has a peak
energy of $7.77$~MeV. The barrier for the $^{12}$C$+$ $^{24}$O system occurs at a
slightly larger $R$ value with a barrier peak of $6.64$~MeV.
Figure~\ref{fig1}b shows the corresponding cross sections for the two reactions.
Also shown are the experimental data from Ref.~\cite{c12o16exp}. The
DC-TDHF potential reproduces the experimental cross-sections quite well for the
$^{12}$C$+$ $^{16}$O system, and the cross section for the neutron rich $^{12}$C+$^{24}$O
is predicted to be larger than that for $^{12}$C +$^{16}$O.

Figures~2a and Fig.~2b show the corresponding potentials and cross-sections for the Ca$+$Ca system~\cite{DC-TDHF-Ca},
which was the subject of recent experimental studies~\cite{Mon11}.
The observed trend for sub-barrier
energies is typical for DC-TDHF calculations when the underlying microscopic interaction
gives a good representation of the participating nuclei. Namely, the potential barrier
corresponding to the lowest collision energy gives the best fit to the sub-barrier cross-sections
since this is the one that allows for more rearrangements to take place and grows the inner part
of the barrier. Considering the fact that historically the low-energy sub-barrier cross-sections
of the $^{40}$Ca+$^{48}$Ca system have been the ones not reproduced well by the standard models,
the DC-TDHF results are quite satisfactory, indicating that the dynamical evolution of the
nuclear density in TDHF gives a good overall description of the collision process.
The shift of the cross-section curve with increasing collision energy is typical.
In principle one could perform a DC-TDHF calculation at each energy above the barrier
and use that cross-section for that energy. However, this would make the computations
extremely time consuming and may not provide much more insight.
The trend at higher energies for the $^{40}$Ca+$^{48}$Ca system is atypical. The calculated cross-sections
are larger than the experimental ones by about a factor of two.
Such lowering of fusion cross-sections with increasing collision energy
is commonly seen in lighter systems where various inelastic channels,
clustering, and molecular formations are believed to be the contributing
factors.
\begin{figure}[!htb]
\includegraphics*[height=.28\textheight]{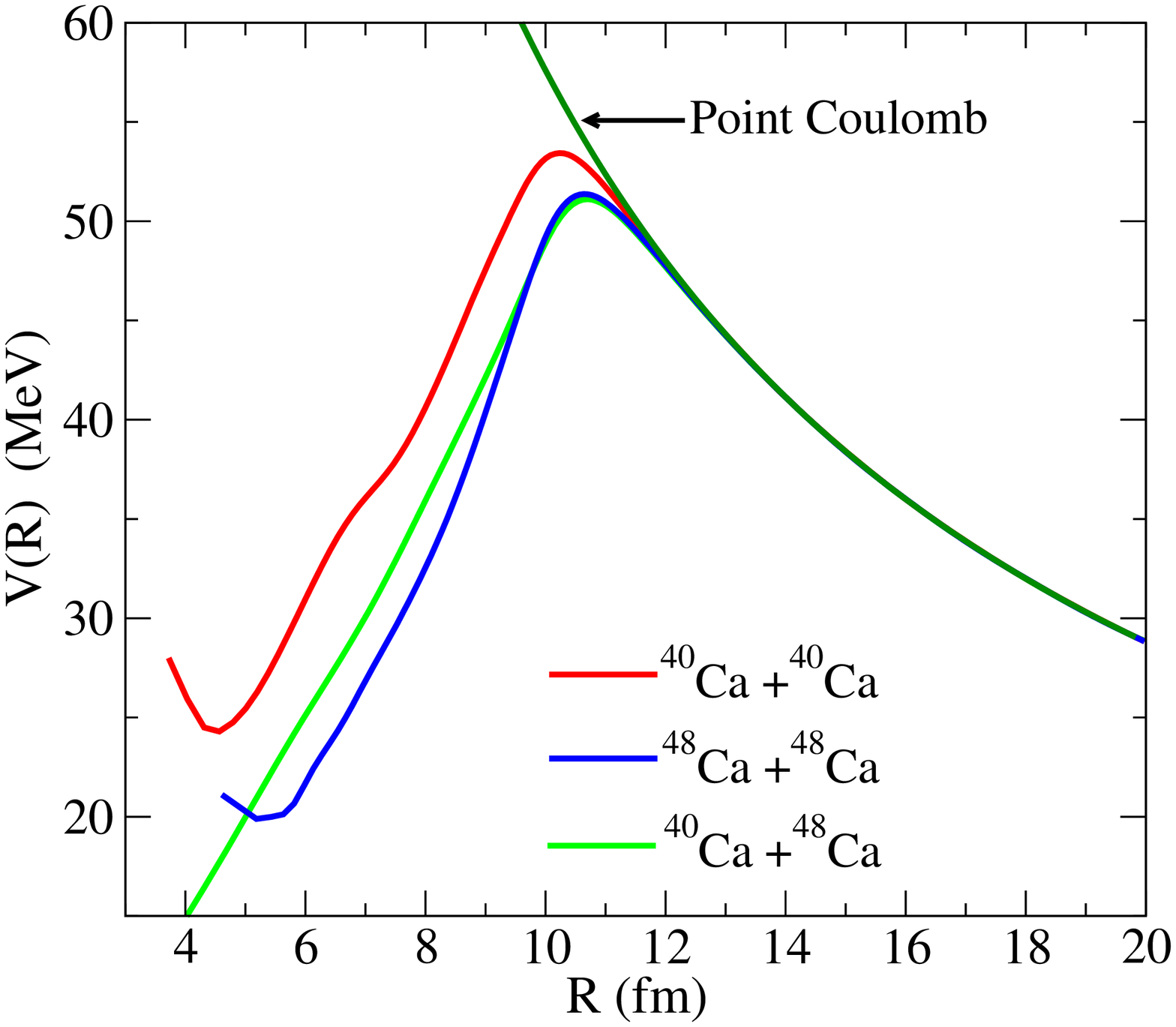}\hspace{0.04in}\includegraphics*[height=.28\textheight]{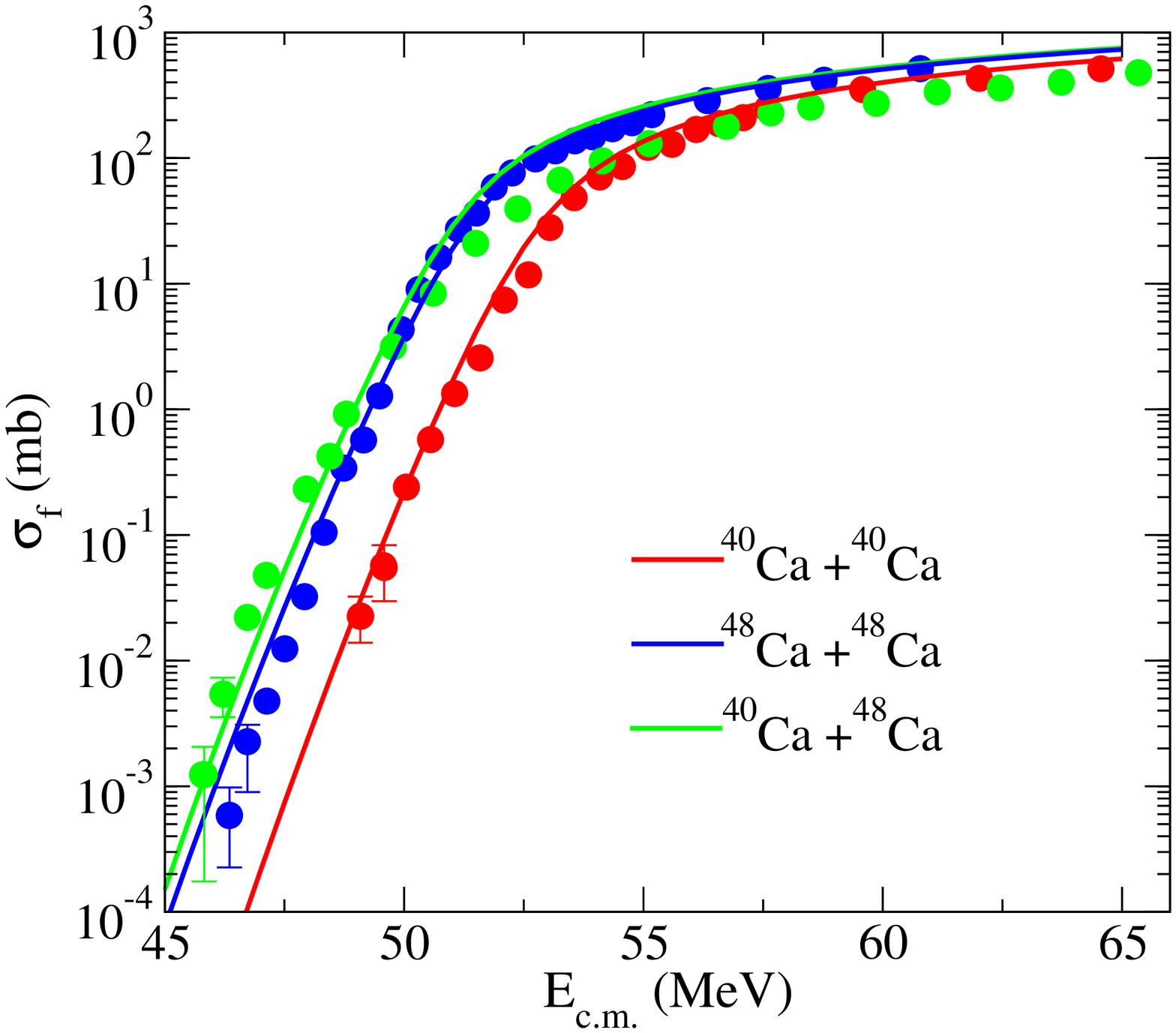}
\caption{\label{fig2} (a) Ion-Ion potential for various isotopes of the Ca$+$Ca system.
                      (b) Corresponding cross-sections.}
\end{figure}

\section{Conclusions}
The fully microscopic TDHF theory has shown itself to be rich in
nuclear phenomena and continues to stimulate our understanding of nuclear dynamics.
The time-dependent mean-field studies seem to show that the dynamic evolution
builds up correlations that are not present in the static theory.
While modern Skyrme forces provide a much better description of static nuclear properties
in comparison to the earlier parametrizations there is a need to obtain even better
parametrizations that incorporate deformation and reaction data into the fit process.

\begin{theacknowledgments}
This work has been supported by the U.S. Department of Energy under grant No.
DE-FG02-96ER40975 with Vanderbilt University, and by the German BMBF
contract Nos. 06FY9086 and 06ER142D.
\end{theacknowledgments}

\end{document}